\documentclass[extra]{gji}
\usepackage{timet}

\usepackage{graphicx}

\title[Saturation and time dependence of geodynamo models]{Saturation and time dependence of geodynamo models}
\author[M. Schrinner et al.]{M. Schrinner$^{1,3}$\thanks{E-mail: martin@schrinner.eu}, D. Schmitt$^{1}$, R. Cameron$^{1}$ and P. Hoyng$^{2}$\\
$^{1}$Max-Planck-Institut f\"ur Sonnensystemforschung, 37191 Katlenburg-Lindau, Germany\\
$^{2}$SRON Netherlands Institute for Space Research, 3584 CA Utrecht, The Netherlands\\
$^{3}$Ecole Normale Sup\'erieur, LRA D\'epartement de Physique, 75005 Paris, France}

\begin{document}

\date{Accepted Received}

\pagerange{\pageref{firstpage}--\pageref{lastpage}} \pubyear{2009}

\maketitle

\label{firstpage}

\begin{abstract}
In this study we address the question under which conditions a saturated
velocity field stemming from geodynamo simulations leads to an exponential
growth of the magnetic field in a corresponding kinematic calculation. We
perform global self-consistent geodynamo simulations and calculate the
evolution of a kinematically advanced tracer field. The self-consistent
velocity field enters the induction equation in each time step, but the tracer
field does not contribute to the Lorentz force. This experiment has been
performed by \cite{cattaneo09} and is closely related to the test field method
by Schrinner et al. (2005, 2007). We find two dynamo regimes in which the
tracer field either grows exponentially or approaches a state aligned with the
actual self-consistent magnetic field after an initial  transition period. Both
regimes can be distinguished by the Rossby number and coincide with the dipolar
and multipolar dynamo regimes identified by \cite{christensen06}. Dipolar
dynamos with low Rossby number are kinematically stable whereas the tracer
field grows exponentially in the multipolar dynamo regime. This difference in
the saturation process for dynamos in both regimes comes along with differences
in their time variability. Within our sample of 20 models, solely kinematically
unstable dynamos show dipole reversals and large excursions. The complicated
time behaviour of these dynamos presumably relates to the alternating growth of
several competing dynamo modes. On the other hand, dynamos in the low Rossby
number regime exhibit a rather simple time dependence and their saturation
merely results in a fluctuation of the fundamental dynamo mode about its
critical state.
\end{abstract}


\section{Introduction}

The time variability of cosmic magnetic fields has always been an argument in
favour of hydromagnetic dynamo action. Its understanding is crucial for
insights in the interior dynamics of stars and planets. The time dependence of
convective dynamos is attributable to a non-stationary buoyancy flux as well as
to a time dependent equilibration of the magnetic field. The latter is subject
of the study presented here.

How do dynamos saturate and in particular in which way is the saturation
reflected in their time dependence? In a general description, the infinite
growth of a magnetic field due to an appropriate motion of a conducting fluid
is inhibited owing to the backreaction of the Lorentz force on the flow; the
resulting changes in the flow cause a  reduction of dynamo action. Flows which
are influenced by the Lorentz force in this way are called saturated.
Nevertheless, \cite{cattaneo09} as well as \cite{tilgner08} demonstrate that
saturated flows may lead to exponential growth of the magnetic field in a
corresponding kinematic calculation. Despite the fact that the magnetic field
is saturated in the full non-linear system, it can grow in a kinematic
treatment, because both associated linearized stability problems are
different. The flows taken from a saturated dynamo simulation and then used in
a kinematic calculation need only quench the growth of the particular magnetic
field found in the nonlinear problem and can in principle allow others to grow.
As \cite{tilgner08} have pointed out there is at least one example, the
benchmark dynamo case~1 \citep{christensen01}, where the field taken from a
saturated dynamo is also kinematically stable.

In this study, we show that there is in fact a whole class of saturated, 
chaotic, time-dependent dynamos for which the corresponding kinematic dynamo 
is stable.
In order to assess  kinematic stability -- in the sense explained above -- we
solve the MHD-equations for a Boussinesq fluid in a rotating spherical shell.
At the same time we evolve a second passive tracer field using the induction
equation. While the tracer field experiences the self-consistent velocity field
at each time step it does not contribute to the Lorentz force. This method has
been used by \cite{cattaneo09} for box simulations and a shell model and is
closely related to the test-field method to determine mean-field coefficients
\citep{schrinner05,schrinner07}.

Within a sample of 20 models, we identify two distinct dynamo regimes dependent
on a modified Rossby number \citep{christensen06} in which the tracer field
either grows exponentially  or reaches a state aligned with the actual
self-consistent magnetic field after an initial transition period. Moreover,
differences in the  kinematic stability of the  dynamos are linked to
differences in their time variability: Exclusively kinematically unstable
dynamos in the high Rossby number regime show polarity reversals of the axial
dipole field. We attribute the complicated time behaviour of these models to an
alternating growth of many competing dynamo modes. On the other hand, the
eigenvalue computation suggests that dynamos with low Rossby number are
dominated by only one fundamental mode which is repeatedly quenched and
rebuilt. All other modes in this case are clearly subcritical. In this sense,
dynamo models in the low Rossby number regime, i.e. fast rotators, exhibit a
simple time dependence and their time-variability consists of fluctuations
about their critical state.

\section{Dynamo calculations}

We consider an electrically conducting Boussinesq fluid in a rotating spherical
shell and solve the MHD-equations as given by \cite{olson99} and described in
detail by \cite{christensen07}. In addition, we compute the evolution of a
passive tracer field with the help of a second induction equation
\begin{equation}
\partial\mitbf{B}_\mathrm{Tr}/\partial t=\mitbf{\nabla}\times(\mitbf{u}
\times\mitbf{B}_\mathrm{Tr})+1/Pm\nabla^2\mitbf{B}_\mathrm{Tr}
\end{equation}
While the tracer field, \(\mitbf{B}_\mathrm{Tr}\), experiences the
self-consistent velocity field \(\mitbf{u}\) in each time step, it does not
contribute to the Lorentz force. Hence it does not act on the velocity field
and is ``passive'' in this sense. The initial conditions for the tracer field
have been chosen arbitrarily with the help of a random number generator.
Moreover, for models 10--15, we added some random noise to the tracer field in
each time step. This enables us to perturb the tracer field permanently and
prevents it from becoming aligned with the actual, self-consistent field. In
these simulations, we advance the tracer field for at least 10 magnetic
diffusion times in order to test for kinematic stability.

According to the scaling we used, the equations are governed by four
parameters. These are the Ekman number \(E=\nu/\Omega D^2\), the (modified)
Rayleigh number \(Ra=\alpha_T g_0\Delta T D/\nu\Omega\), the Prandtl number
\(Pr=\nu/\kappa\) and the magnetic Prandtl number \(Pm=\nu/\eta\). In these
expressions,  \(\nu\) denotes the kinematic viscosity, \(\Omega\) the rotation
rate, \(D\) the shell width, \(\alpha_T\) the thermal expansion coefficient,
\(g_0\) is the gravitational acceleration at the outer boundary, \(\Delta T\)
stands for the temperature difference between the inner and outer spherical
boundaries, \(\kappa\) is the thermal and \(\eta=1/\mu\sigma\) the magnetic
diffusivity with the magnetic permeability \(\mu\) and the electrical
conductivity \(\sigma\). All four parameters have been varied to build up a
sample of 20 dynamo models, see Table~1.

Output parameters used here in order to interpret the results are the magnetic
Reynolds number, \(Rm=UD/\eta\), the Elsasser number,
\(\Lambda=B^2/\varrho\mu\eta\Omega\), and the Rossby number, \(Ro=U/D\Omega\).
In these expressions, \(U\) and \(B\) denote rms-values of the velocity and the
magnetic field inside the shell, respectively, and \(\rho\) is the density.
Furthermore, we adopt the definition of a local Rossby number proposed by
\cite{christensen06},
\begin{equation}
Ro_l=Ro\cdot\frac{\bar{l}}{\pi}
\end{equation}
 Here, \(\bar{l}/{\pi}\) is the mean half wavelength of the flow and
\(\bar{l}\) is the mean harmonic degree derived from the kinetic energy
spectrum,
\begin{equation}
\bar{l}=\sum_ll\frac{<\mitbf{u}_l\cdot\mitbf{u}_l>}{<\mitbf{u}\cdot\mitbf{u}>}
\label{eq:3}
\end{equation}
The brackets in Eq.~(\ref{eq:3}) denote an average over time and radii,
\(\mitbf{u}_l\) stands for the velocity component of harmonic degree \(l\).

\section{Results}

\begin{figure}
\includegraphics{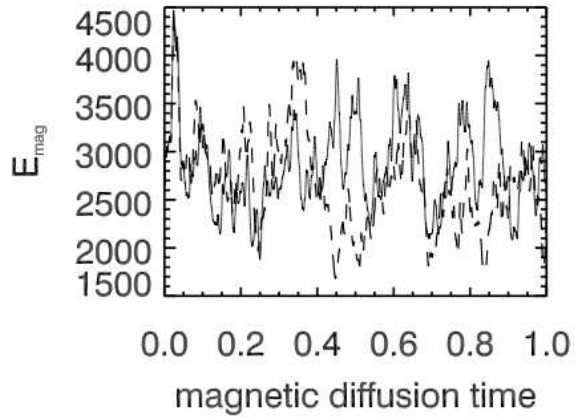}
\caption{Magnetic energy densities for two computational runs of model 2. Both
runs have been started from very similar initial conditions which differ only
by a small deflection (dashed line) of the magnetic dipole axis. Nevertheless,
both models evolve differently which demonstrates the chaotic character of
these dynamos.}
\label{figure7}
\end{figure}

\begin{figure}
\includegraphics{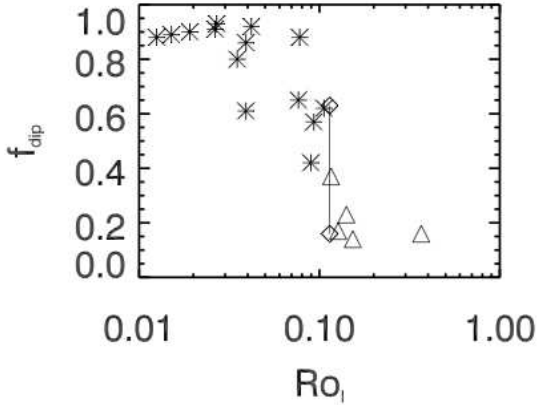}
\caption{Relative dipole field strength \(f_\mathrm{dip}\) versus local Rossby
number \(Ro_l\). Stars denote non-reversing dynamos which are kinematically
stable, whereas triangles represent dynamos which do reverse and are
kinematically unstable. Both regimes coincide with the dipolar and multipolar
dynamo regimes identified by Christensen and Aubert (2006). There is one
example (diamonds), model 15, which undergoes a transition between both
regimes. Note that this example has a considerably lower relative dipole field
strength in its second state.}
\label{figure1}
\end{figure}

Within our 20 examples (see Table \ref{tab1}) we find 5 dynamos which are
kinematically unstable and 14 which are kinematically stable. One example
(model 15)  belongs to both classes; although in general unstable, the tracer
field does not grow within certain periods of several magnetic diffusion times.
Note that all dynamos considered here operate in the so called strong field
regime, i.e. the Elsasser number is of order unity or larger. The equatorial
symmetry is broken for most of the kinematically stable and all unstable
models. Except for model 1, the quasi-steady benchmark dynamo
\citep{christensen01}, all models exhibit highly time-dependent or even chaotic
behaviour. This is demonstrated in Fig.~\ref{figure7} for model 2, the next
simplest example to the benchmark dynamo. This dynamo appears to be chaotic,
and as an experiment we performed two simulations starting from almost
identical initial conditions (the difference between two initial conditions is
a small deflection of the magnetic dipole axis in the second run). The
evolution from both initial conditions is shown in Fig.~\ref{figure7} where
the magnetic energy densities can be seen to diverge rapidly with time.

\begin{figure}
\includegraphics{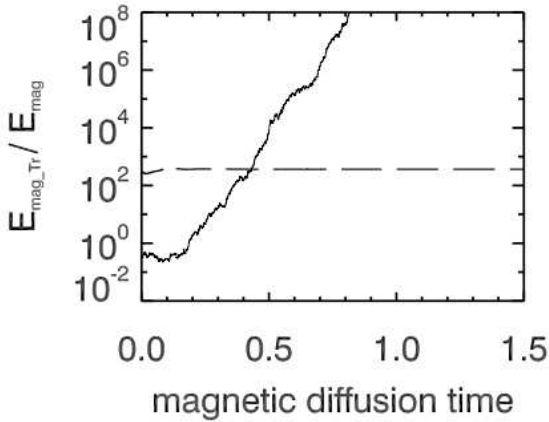}
\caption{Ratio of the magnetic energy densities for the tracer field,
\(E_\mathrm{mag_{Tr}}\), and the actual magnetic field, \(E_\mathrm{mag}\), versus
time for model 8 (dashed line) and model 19 (solid line).}
\label{figure4}
\end{figure}

\begin{figure}
\includegraphics{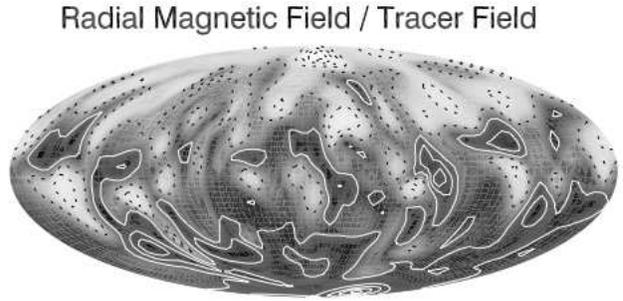}
\caption{Snapshot of the radial component of the actual magnetic field and the
tracer field for model 8 taken some time after an initial transition period
at \(r=0.62r_o\) where \(r_o\) is the outer shell radius. Note that the
tracer field is completely aligned with the actual magnetic field. Both
components are normalised due to their maxima and minima. Therefore the
greyscale coding varies from -1, white, to +1, black, and the contour lines
correspond to \(\pm 0.1,\pm 0.3, \pm 0.5, \pm 0.7, \pm 0.9\). Following contour
plots are presented in the same style.}
\label{figure2}
\end{figure}

\begin{figure}
\includegraphics{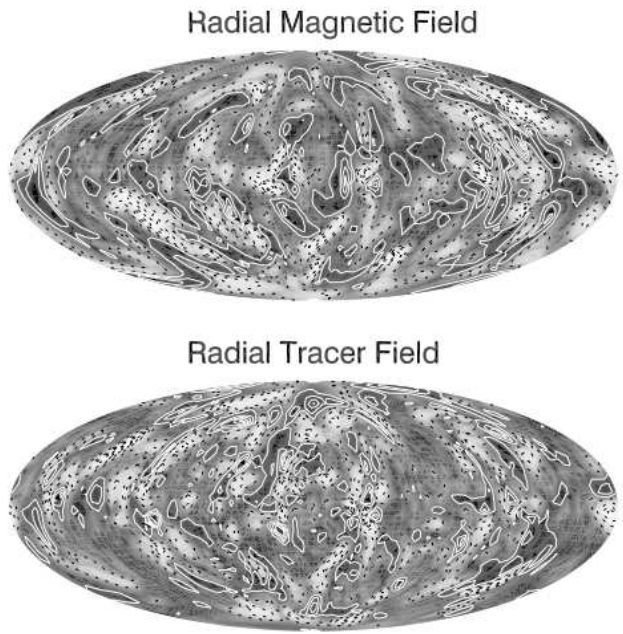}
\caption{Snapshots of the radial component of the actual magnetic field
(top) and the tracer field (bottom) for model 19 at \(r=0.62r_o\). Contour
lines: see figure \ref{figure2}.}
\label{figure3}
\end{figure}

The regimes of kinematically stable and unstable dynamos can be clearly
distinguished by the modified Rossby number (see Table \ref{tab1}), \(Ro_l\).
Models with low Rossby number are kinematically stable whereas the tracer field
grows exponentially for dynamos in the high Rossby number regime. The
transition between both regimes occurs at \(Ro_l\approx 0.12\). There are two
further properties related to both regimes which deserve mentioning. All
dynamos we found to be kinematically stable are dipolar and do not show any
polarity reversals, while dynamos in the second regime are multipolar and do
reverse. This is also illustrated in Fig.~\ref{figure1}. Here, the relative
dipole field strength, \(f_\mathrm{dip}\), on the outer shell boundary is plotted
versus the modified Rossby number, \(Ro_l\); \(f_\mathrm{dip}\) is defined as the
time-average ratio of the dipole field strength to the field strength in
harmonic degrees 1 to 12. Both regimes visible in Fig.~\ref{figure1} coincide
with those identified earlier by \cite{christensen06}. Figure~\ref{figure4}
compares the magnetic energy densities of the tracer field for a kinematically
stable (model 8) and a kinematically unstable dynamo (model 19), varying with
time. While the tracer field grows rapidly after an initial transient phase in
the latter case, it reaches a state aligned with the actual field if the dynamo
is kinematically stable. Then, the  energy density of the tracer field
normalised with the energy density of the actual self-consistent field
approaches a constant level which depends only on the initial conditions. This
is also confirmed by looking at the corresponding field configurations.
Figure~\ref{figure2} displays the radial component of the tracer field for model 8,
which differs from the actual field only by an overall scale factor. Therefore,
only one contour plot is given. On the other hand, although they have
similar spatial scales, both field components are clearly not aligned but very
different for model 19 (see Fig.~\ref{figure3}).

\begin{table*}
  \begin{minipage}{140mm}
    \caption{Overview of the runs considered, ordered with respect to their
     modified Rossby number. All kinematically unstable models exhibit dipole
     reversals whereas all kinematically stable models do not.}
    \label{tab1}
    \centering
    \begin{tabular}{@{}lcrrrcrlrrr@{}}
      \hline
      Model & $E$ & $Ra$ & $Pm$ & $Pr$ & $Ro$ & mean \(l\) & \(Ro_l\)&
      \(f_\mathrm{dip}\) & $Rm$ &\(\Lambda\)\\
      \hline
      &&&&&&&&&&\\

      &&&\multicolumn{4}{c}{Kinematically stable models}\\
      &&&&&&&&&&\\
 model 1  & \(1\times 10^{-3}\) & 100& 5 & 1 & 0.0079 & 5  & 0.013 & 0.88&39&6.3 \\
 model 2  & \(1\times 10^{-4}\) & 334& 2 & 1 & 0.0043 & 11 & 0.015 & 0.89&86&1.0 \\
 model 3  & \(3\times 10^{-4}\) & 195& 3 & 1 & 0.0067 & 9 & 0.019 & 0.92&67&0.6 \\
 model 4  & \(3\times 10^{-4}\) & 243& 2 & 1 & 0.0085 & 9 & 0.024 & 0.93&56&1.7 \\
 model 5  & \(3\times 10^{-4}\) & 285& 2 & 1 & 0.0092 & 9 & 0.026 & 0.91&61&2.2 \\
 model 6  & \(3\times 10^{-4}\) & 375& 3 & 1 & 0.0110 & 10 & 0.035 & 0.80&110&5.7 \\
 model 7  & \(3\times 10^{-4}\) & 330& 9 & 3 & 0.0094 & 13 & 0.039 & 0.63&283&11.9 \\
 model 8  & \(3\times 10^{-4}\) & 330& 3 & 3 & 0.0094 & 13 & 0.039 & 0.86&95&2.7 \\
 model 9  & \(3\times 10^{-4}\) & 375& 1.5 & 1 & 0.0120 & 11 & 0.042 & 0.92&60&2.0 \\
 model 10 & \(3\times 10^{-4}\) & 630& 3 & 1 & 0.0200 & 12 & 0.076 & 0.65&200&6.8 \\
 model 11 & \(1\times 10^{-4}\) & 1117& 1.5 & 1 & 0.0128 & 19 & 0.078 & 0.88&129&2.3 \\
 model 12 & \(1\times 10^{-3}\) & 400& 10 & 1 & 0.0352 & 8 & 0.090 & 0.42&352&20.0 \\
 model 13 & \(3\times 10^{-4}\) & 810& 5 & 1 & 0.0244 & 12 & 0.093 & 0.57&406&18.0 \\
 model 14 & \(3\times 10^{-4}\) & 750& 3 & 1 & 0.0257 & 13 & 0.106 & 0.62&257&5.5 \\
&&&&&&&&&&\\
 &&&\multicolumn{4}{c}{Kinematically unstable models}\\
&&&&&&&&&&\\
 model 15 & \(3\times 10^{-4}\) & 810& 3 & 1 & 0.0276 & 13 & 0.114 & 0.61 (0.16)&276&4.7 \\
 model 16 & \(1\times 10^{-3}\) & 450& 10& 1 & 0.0406 & 9 & 0.116  & 0.37&406&19.0 \\
 model 17 & \(1\times 10^{-3}\) & 500& 10& 1 & 0.0442 & 9 & 0.127  & 0.17&442&10.5 \\
 model 18 & \(3\times 10^{-4}\) & 1050& 3& 1 & 0.0340 & 13 & 0.141  & 0.23&341&2.2 \\
 model 19 & \(3\times 10^{-4}\) & 1250& 3& 0.3 & 0.0479 & 10 & 0.153  & 0.14&479&7 \\
 model 20 & \(3\times 10^{-4}\) & 2970& 1& 0.3 & 0.1154 & 10 & 0.367 & 0.16&385&0.4 \\
\hline
\end{tabular}
\end{minipage}
\end{table*}

Model 15 is in general kinematically unstable but also exhibits periods of
several magnetic diffusion times in which the tracer field stays stable.
According to its local Rossby number, \(Ro_l=0.114\), it is located close to
the boundry between both dynamo regimes and undergoes transitions from one to
the other.

\begin{figure}
\includegraphics{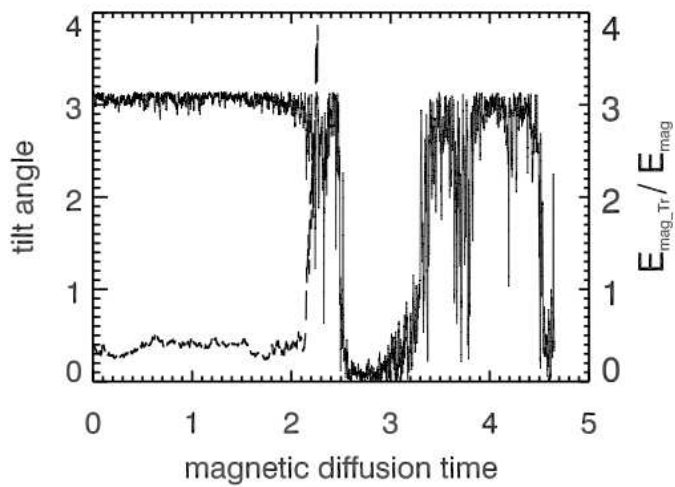}
\caption{Tilt angle of the dipole axis for model 15 as a function of time 
(solid line) and magnetic energy density of the tracer field normalised by the
magnetic energy density of the actual field,
\(E_\mathrm{mag_{Tr}}/E_\mathrm{mag}\) (dashed line which runs out of the 
figure at roughly 2.3 magnetic diffusion times). As soon as the dynamo
reverses it becomes kinematically unstable.}
\label{figure4a}
\end{figure}

We could detect transitions from a kinematically stable to an unstable state
(see Fig.~\ref{figure4a}) and vice versa. As long as the tracer field remains
stable, the tilt angle of the dipole axis fluctuates about the actual polarity
state. However, when the tracer field becomes unstable, also the polarity of
the dipole field starts to reverse. This coincidence is observed for
transitions in both directions, i.e. the tilt angle of the dipole axis also
stabilises when model 15 becomes intermittently stable. While the magnetic
field is quite dipolar with \(f_{\mathrm{dip}}=0.61\) for periods in which the
polarity and the tracer field are stable the relative dipole field strength
decreases drastically to \(f_{\mathrm{dip}}=0.16\) otherwise. The strong
connection of field morphology, time dependence and saturation is not only
present separately in several models but manifests itself in the time variation
of a single dynamo model, too.

\section{Discussion}

The existence of kinematically unstable dynamos was expected
\citep{cattaneo09,tilgner08}. The finding of a class of kinematically stable
but yet time-dependent or even chaotic dynamos, however, needs some further
explanation. The lack of growing modes for these models already suggests that
almost all field configurations for the tracer field are decaying, except the
one aligned with the actual, self-consistent field. But this component of the
tracer field is quenched by the saturated velocity field. Thus, the tracer
field follows the actual field with time, apart from a different, arbitrary
amplitude due to the linearity of the induction equation.

This interpretation is confirmed by looking at the spectrum of the time and
azimuthally averaged dynamo operator \(D\),
\begin{equation}
D\mitbf{b}^i=\lambda^i\mitbf{b}^i
\end{equation}
with eigenmodes \(\mitbf{b}^i\) and eigenvalues \(\lambda^i\). In this, the
operator  \(D\) is defined as
\begin{equation}
D\mitbf{b}=\mitbf{\nabla}\times(\mitbf{\bar{u}}\times\mitbf{b}+\mitbf{\alpha}\cdot\mitbf{b}
-\mitbf{\beta}\,\mitbf{\nabla}\mitbf{b}-\eta\mitbf{\nabla}\times\mitbf{b})
\end{equation}
Note that \(D\), also known as mean-field dynamo operator \citep{raedler},
contains the mean velocity field \(\mitbf{\bar{u}}\) as well as the so called
mean-field coefficients \(\mitbf{\alpha}\) and \(\mitbf{\beta}\), which are
tensors of second and third rank, respectively.  As noted by \citet{hoyng09},
these quantities appear inevitably as a consequence of averaging. They depend
on the velocity field and the magnetic diffusivity of the considered dynamo
model only and have been determined with the help of the test field method
\citep{schrinner05,schrinner07}. A detailed discussion on the applicability of
mean-field concepts to direct numerical simulations of rotating
magnetoconvection and a (quasi-)stationary dynamo is provided by
\cite{schrinner07}. A similar discussion for time dependent dynamos is not
given here but will be subject of a forthcoming paper. A recent review on the
test-field method and its applications has been given by \cite{brandenburg09}.

\begin{figure}
\includegraphics{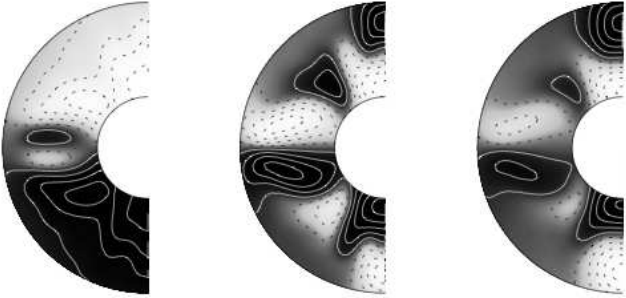}
\caption{Radial components of the thirst three dipolar eigenmodes
  \(\mitbf{b}^i,i=1\dots 3\) of the time averaged dynamo operator for
  model 2. The corresponding eigenvalues are
  \(\lambda_1=-3.87,\,\lambda_2=-34.83\) and
  \(\lambda_3=-42.45\) in units of \(\eta/D^2\). Note the huge drop in decay
  rates after \(\lambda_1\).}
\label{figure5}
\end{figure}

Eigenvalues and eigenmodes of \(D\) have been computed as reported by
\cite{schrinner09} for model 2. In Fig.~\ref{figure5} the radial components of
the first three (dipolar) eigenmodes, \(\mitbf{b}^i\), are displayed. All modes
decay exponentially; this had to be expected for kinematically stable dynamos
\citep[see also the discussion in][]{hoyng09}. However, the decay rates are
given here in units of \(\eta/D^2\), in which the molecular diffusivity
\(\eta\) is about 30 times smaller than the turbulent one inferred from
components  of \(\mitbf{\beta}\). Thus, \(1/|\lambda_1|\approx 1/4\) is much
larger than one  effective diffusion time and the first, fundamental, eigenmode
is indeed close to its critical state. Due to a noticable gap in decay  rates
after the fundamental mode, $|\lambda_1| \ll |\lambda_2|$, this is not equally
true for the subsequent eigenmodes. They are much more diffusive, thus  leaving
the fundamental mode as the preferred eigenstate of the dynamo. Hence, the time
dependence of model 2 may be understood in parts as a fluctuation of the
fundamental mode about its critical state.

The dominance of the first eigenstate is also revealed by a
decomposition of the actual, time-dependent magnetic field of model 2 in a set 
of eigenmodes \(\mitbf{b}^i\) of \(D\),
\begin{equation}
\mitbf{B}(\mitbf{r},t)=\sum_ia^i(t)\mitbf{b}^i(\mitbf{r})
\label{eq:7a}
\end{equation}
The time-dependent and in general complex mode coefficients \(a^i(t)\) have
been computed as
\begin{equation}
a^i(t)=\int_V\mitbf{\hat{j}}^{i}(\mitbf{r})\cdot\mitbf{A}(\mitbf{r},t) \,\mathrm{d^3}\mitbf{r}
\label{eq:7}
\end{equation}
in which \(\mitbf{\hat{j}}^i\) denotes the adjoint of the current
\(\mitbf{j}^i=\mitbf{\nabla}\times\mitbf{b}^i\), and
\(\mitbf{A}\) is the vector potential of the actual,
time-dependent field, \(\mitbf{B}=\mitbf{\nabla}\times\mitbf{A}\). The integration
is carried out over the whole fluid domain \(V\). For a derivation of Eq.~(\ref{eq:7}) we
refer to \citet{hoyng09} and \citet{schrinner09}. 

\begin{figure}
\includegraphics{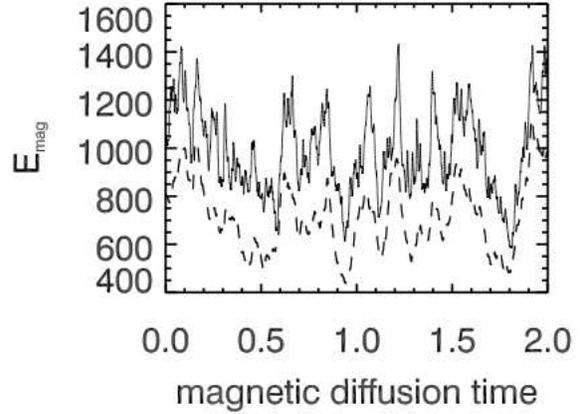}
\caption{Axisymmetric magnetic energy density (solid line) and energy
contribution of the first, fundamental eigenmode (dashed line) varying with
time. The fundamental eigenmode \(a^1(t)\mitbf{b}^1(\mitbf{r})\) already
contributes 75\% up to 85\% to the total amount and its time variability
reflects much of the time dependence of the axisymmetric magnetic field.}
\label{figure6}
\end{figure}

In Fig.~\ref{figure6} the energy contribution of the fundamental eigenmode
\(a^1(t)\mitbf{b}^1(\mitbf{r})\) is compared with the total axisymmetric
magnetic energy density. The fundamental eigenmode contributes at least 75\% up
to 85\% to the total amount, revealing again its permanent dominance throughout
the simulation.

The equilibration process for model 2 has been studied earlier 
by \cite{olson99}, too. They found that in regions with high magnetic energy 
density, the Lorentz force simply reduces locally the flow velocity without 
changing the overall pattern of convection. They investigated possible changes 
in the velocity, if
the magnetic field and thus the Lorentz force is arbitrarily reduced at some
instant in time and then recovers towards its equilibration value. The
kinematic effects relevant for dynamo action identified by them, an
\(\alpha\)-effect from helicity in the columnar convection and an
anti-\(\omega\) effect from the mean azimuthal flow, were present in the same
proportions close and far from equilibrium conditions of the magnetic field.
Their finding is supported by the study presented here. Saturation may reduce
the amplitudes of \(\alpha\) and thus the growth rates of the eigenstates of
the related dynamo operator, but does not change their relative order.
Therefore, the preferred eigenstate stays the same throughout the simulation.
This is clear as there is a large gap between the growth rate of the 
fundamental eigenmode of the time-averaged dynamo operator \(D\) and all other 
eigenmodes, as mentioned above.

So far we only analysed model 2 in detail. Here the velocity field is nearly
symmetric with respect to the equatorial plane and the magnetic field 
belongs to the dipolar family. Contributions of quadrupolar type are not
present. In a more complicated example with broken equatorial symmetry, we
expect the fundamental quadrupolar mode to be excited. Although its growth rate
will be smaller than the one for the fundamental dipolar mode and probably
subcritical, it is typically of the same order. In such a case a clear
dominance of only one fundamental dipolar mode can no longer be deduced from 
the spectrum of the time-averaged dynamo operator \(D\), and a second, 
quadrupolar mode may become important.

For the models in the high Rossby number regime the findings of
\cite{cattaneo09} apply. These models act kinematically as dynamos and the
dynamo operator \(D\) possesses in general growing eigenmodes. A kinematic
treatment of these dynamos does not reveal their actual time dependence.
However, the results presented here suggest that the regime of dipolar dynamos
identified by \cite{christensen06}  is kinematically stable. For these models,
the quenching of any magnetic field is fully captured in the velocity field and
a kinematic treatment may indeed reproduce their actual time-dependence. Models
of this dynamical regime are applicable to planetary dynamos and probably also
to dynamos of fast rotating stars \citep{christensen09a}, thus covering a large
range of magnetic Reynolds numbers. Hence, an attempt to explain the kinematic
stability of these models due to a magnetic Reynolds number which is close to
its critical value fails. We emphasise again that the  transition between both
regimes is governed by the local Rossby number \(Ro_l\)  and not by \(Rm\), as
can be already seen from Table~\ref{tab1}. In the low Rossby number regime, the
rotational constraint leads to columnar structured flows, dipolar magnetic
fields and finally to a rather simple time dependence, although these models
operate in general far away from the dynamo threshold at \(Rm_c\approx 40\).

Dipolar dynamo models which show occasionally reversals are located close to
the regime boundary in Fig.~\ref{figure1}, with \(Ro_l\le 0.12\). They
resemble the geodynamo in many respects and are therefore of particular
interest. Explaining polarity reversals of an otherwise predominantly dipolar
field, \cite{olson06} suggest that the geodynamo crosses the boundary towards
the multipolar dynamo regime from time to time. With the help of scaling laws
derived from numerical models, they indeed succeed in predicting a local Rossby
number of \(Ro_l\approx 0.09\) for the Earth's core. Adopting this viewpoint we
link the occurence of geomagnetic reversals to a change in the saturation
process. The quenching of a previously dipolar field may result in the
preference of different, higher order modes if inertia gains importance in
comparison to the coriolis force, and the dynamo undergoes an excursion into
the kinematically unstable regime. Subsequently the dipole field is built up
again, but it may have either polarity. A computation of eigenmodes and a  mode
decomposition similar to (\ref{eq:7a}) for model 15 seems to be a promising
approach to confirm this picture. Note that from the viewpoint we take here,
the existence of dipolar, stable periods for model 15 demands more explanation
than the fact that it reverses.

However, whether inertia is indeed as important for the geodynamo as it is for
present dynamo models is under debate \citep[e.g.][]{sreenivasan06}. In fact,
the assumption of \(Ro_l\approx 0.09\) for the Earth's core leads to a
characteristic length scale of only a few hundred meters, on which the magnetic
field would be highly diffusive \citep{christensen09b}.
%

\section{Conclusions}

Fast rotating dynamos, characterised by a low Rossby number, are kinematically
stable. Within this regime, a saturated velocity field taken from dynamo
simulations does not lead to exponential growth of the magnetic field in a
corresponding kinematic calculation. Hence, saturation may be understood as a
quality of the velocity field, only. For these dynamos, saturation results in
the unchanged preference of a fundamental eigenstate, whereas different
eigenmodes may supersede each other if inertia gains importance. This
difference in the saturation process involves differences in the morphology of
the magnetic field and its time dependence. Kinematically stable dynamos are
dipolar and exhibit a rather simple time variability, which may be interpreted
as the fluctuation of the fundamental mode about its critical state.
Kinematically unstable dynamos are much more complicated. The alternating
growth of various modes leads to a multipolar field morphology and polarity
reversals of the dipole field appear as a natural consequence.

\section*{Acknowledgments}

We thank Ulrich Christensen and Johannes Wicht for many interesting discussions
and support.


\begin{thebibliography}{}
%
\bibitem[Brandenburg(2009)]{brandenburg09}
Brandenburg, A., 2009. Advances in theory and simulations of large-scale dynamos,
\textit{Space Sci. Rev.}, \textbf{144}, 87--104.
%
\bibitem[Cattaneo \& Tobias(2009)]{cattaneo09}
Cattaneo, F. \& Tobias, S.M., 2009. Dynamo properties of the turbulent velocity field
of a saturated dynamo,
\textit{J. Fluid Mech.}, \textbf{621}, 205--214.
%
\bibitem[Christensen et al.(2001)]{christensen01}
Christensen, U.R., Aubert, J., Cardin, P., Dormy, E., Gibbons, S., Glatzmaier, G.A.,
Grote, E., Honkura, Y., Jones, C., Kono, M., Matsushima, M., Sakuraba, A.,
Takahashi, F., Tilgner, A., Wicht, J. \& Zhang, K., 2001.
A numerical dynamo benchmark,
\textit{Phys. Earth. Planet. Inter.}, \textbf{128}, 25--34.
%
\bibitem[Christensen \& Aubert(2006)]{christensen06}
Christensen, U.R. \& Aubert, J., 2006. Scaling properties of convection-driven dynamos
in rotating spherical shells and application to planetary magnetic fields,
\textit{Geophys. J. Int.}, \textbf{166}, 97--114.
%
\bibitem[Christensen \& Wicht(2007)]{christensen07}
Christensen, U.R. \& Wicht, J., 2007. Numerical dynamo simulations, in \textit{Treatise
on Geophysics, Vol. 8}, pp. 245--282, ed. Schubert G., Elsevier, Amsterdam.
%
\bibitem[Christensen et al.(2009a)]{christensen09a}
Christensen, U.R., Holzwarth, V. \& Reiners, A., 2009a.
Energy flux determines magnetic field strength of planets and stars,
\textit{Nature}, \textbf{457}, 167--169.
%
\bibitem[Christensen et al.(2009b)]{christensen09b}
Christensen, U.R., Schmitt, D. \& Rempel, M., 2009b.
Planetary dynamos from a solar perspective,
\textit{Space Sci. Rev.}, \textbf{144}, 105--126.
%
\bibitem[Hoyng(2009)]{hoyng09}
Hoyng, P., 2009.
Statistical dynamo theory: Mode excitation,
\textit{Phys. Rev. E}, \textbf{79}, 046320, 1--13.
%
\bibitem[Krause \& R\"adler(1980)]{raedler}
Krause, F. \& R\"adler, K.-H., 1980.
Mean-Field Magnetohydrodynamics and Dynamo Theory, Pergamon Press, Oxford.
%
\bibitem[Olson \& Christensen(2006)]{olson06}
Olson, P. \& Christensen, U.R., 2006.
Dipole moment scaling for convection driven planetary dynamos,
\textit{Earth Planet. Sci. Lett.}, \textbf{250}, 561--571.
%
\bibitem[Olson et al.(1999)]{olson99}
Olson, P., Christensen, U.R. \& Glatzmaier, G.A., 1999.
Numerical modeling of the geodynamo: Mechanisms of field generation and equilibration,
\textit{J. Geophys. Res.}, \textbf{104}, 10383--10404.
%
%
%
\bibitem[Schrinner et al.(2005)]{schrinner05}
Schrinner, M., R\"adler, K.-H., Schmitt, D., Rheinhardt, M. \& Christensen, U.R., 2005.
Mean-field view on rotating magnetoconvection and a geodynamo model,
\textit{Astron. Nachr.}, \textbf{326}, 245--249.
%
\bibitem[Schrinner et al.(2007)]{schrinner07}
Schrinner, M., R\"adler, K.-H., Schmitt, D., Rheinhardt, M. \& Christensen, U.R., 2007.
Mean-field concept and direct numerical simulations of rotating magnetoconvection and the geodynamo,
\textit{Geophys. Astrophys. Fluid Dynam.}, \textbf{101}, 81--116.
%
\bibitem[Schrinner et al.(2009)]{schrinner09}
Schrinner, M., Schmitt, D., Jiang, J. \& Hoyng, P., 2009.
A new method for computing the eigenfunctions and their adjoints of the dynamo operator,
\textit{Mon. Not. R. Astron. Soc.}, submitted.
%
\bibitem[Sreenivasan \& Jones(2006)]{sreenivasan06}
Sreenivasan, B. \& Jones, C.A., 2006.
The role of inertia in the evolution of spherical dynamos,
\textit{Geophys. J. Int.}, \textbf{164}, 467--476.
%
\bibitem[Tilgner \& Brandenburg(2008)]{tilgner08}
Tilgner, A. \& Brandenburg, A., 2008.
A growing dynamo from a saturated Roberts flow dynamo,
\textit{Mon. Not. R. Astron. Soc.}, \textbf{391}, 1477--1481.
%
\end{thebibliography}
\end{document}